# Lateral Optical Forces on Linearly-Polarized Emitters near a Reciprocal Substrate

Hafssaa Latioui[1, 2], Mário G. Silveirinha[1,2]*

[(1)] *University of Lisbon–Instituto Superior Técnico - Avenida Rovisco Pais, 1, 1049-001 Lisboa, Portugal*

[(2)] *Department of Electrical Engineering, University of Coimbra, and Instituto de Telecomunicações, 3030-290 Coimbra, Portugal*

## Abstract

We theoretically investigate the emergence of lateral (recoil) forces acting on generic dipole-type emitters in the vicinity of a reciprocal translation-invariant substrate. Surprisingly, we find that for linearly-polarized electric dipoles the lateral force invariably vanishes, independent of the anisotropy (e.g., tilted optical axes) or chirality of the substrate. We identify a novel opportunity to have a recoil force relying on a superposition of two linearly-polarized and collinear electric and magnetic dipoles. Counterintuitively, it is shown that when such an emitter stands above a uniaxial dielectric half-space with tilted optical axes it may experience a recoil force oriented along the direction perpendicular to the plane defined by the interface normal and the substrate optical axis.

* To whom correspondence should be addressed: E-mail: mario.silveirinha@co.it.pt

# I. Introduction

The optical manipulation of small particles is important for the laser cooling of neutral atoms, particle transport, optical sorting, just to name a few applications [1]-[7]. The optical trapping operation relies on tightly focused laser beams (optical tweezers), which allow one to precisely position and displace neutral particles using gradient-type optical forces [8-15].

Furthermore, fluctuation-induced optical (Casimir-Polder) forces acting on atoms near the surface of a material are of considerable interest due to their relevance in a variety of processes in physics, biology and chemistry [16]-[17]. In translation-invariant scenarios, e.g., when an atom stands nearby a perfectly smooth material surface, the ground-state Casimir-Polder force is along the direction normal to interface, and typically it pulls the atom towards the surface. Interestingly, despite the translation invariance property, for an excited atom the Casimir-Polder force gains an extra resonant term [18], which can have a lateral component. The lateral force is a consequence of the directional nature of the fields emitted by the excited atom in a spontaneous emission process. Similar lateral forces emerge even when one adopts a purely classical treatment, wherein the atom is modeled as a polarizable particle. In particular, lateral forces have been predicted for chiral particles standing above a metal surface and related platforms [19]-[26] and in a few non-reciprocal systems [27]-[30]. In fact, intuitively one may expect that whenever the radiation pattern of the emitter is asymmetric, let us say when it radiates more energy to the "right" as compared to the "left", then the conservation of momentum should imply the emergence of a lateral (recoil-type) optical force. This is precisely what happens when a chiral emitter loses its energy in a spontaneous emission process above a metallic surface [20]-[22]. In this case, the atom will couple selectively to plasmons with a certain handedness, and hence due to the spin momentum locking property [31], it will emit more strongly towards certain directions of space.

Motivated by this result, one may wonder if linearly polarized dipoles standing above a reciprocal material half-space can experience similar lateral forces. For example, heuristically one may expect the emergence of lateral forces when an atom or classical particle relaxes from an excited state to the ground near a standard uniaxial material half-space with the optical axis tilted with respect to the interface (see Ref. [32] for a related problem). Furthermore, since an excited chiral-particle above a dielectric substrate can experience a non-trivial lateral force [20]-[21], one may wonder if the chiral property can be "swapped" with the substrate, i.e., if an excited linearly polarized particle above a chiral substrate may experience a nontrivial recoil-force. Surprisingly, we prove here that for a linearly-polarized particle with a purely electric response (electric dipole approximation) the lateral force vanishes for any reciprocal translation-invariant substrate, independent of the orientation of the optical axes or of the bianisotropy (chirality) of the involved materials. In particular, it follows that the chiral response of the particle cannot be "transferred" to the substrate. Interestingly, we propose a new mechanism to have non-trivial lateral forces with linearly-polarized particles with nontrivial electric and magnetic responses. For the sake of simplicity, we adopt a classical picture with the particle regarded as a superposition of electric and magnetic dipoles.

# II. Optical Force on a Single Electric Dipole

To begin with, it is supposed that the particle has a purely electric response so that it can be approximated by an electric dipole. It is assumed that the dipole oscillations are either pumped by a gradient-less external excitation, or, alternatively, that the oscillations are due to the relaxation of some natural mode with a complex-valued resonant frequency $\omega = \omega_0' - i\omega_0''$, with $\omega_0'' > 0$. The decay rate of the dipole oscillations is determined by $\omega_0''$.

## *A. Lateral force on a linearly polarized dipole*

Let us suppose that a generic particle with electric dipole moment $\mathbf{p}_e$ stands in air near the flat surface of an arbitrary smooth translation-invariant reciprocal substrate. The direction normal to the substrate is along *z* (see a generic geometry in Fig. 1a). For now, we consider the scenario wherein the dipole oscillations are driven by an incident beam so that the oscillation frequency $\omega$ is real-valued. The case wherein the dipole oscillations are damped ($\omega$ is complex-valued) and the emitted radiation is due to the decay from an excited state is discussed in Sect. II.D. The time-averaged optical force acting on the electric dipole for a time harmonic oscillation is (the time variation $e^{-i\omega t}$ is implicit),

$$\mathscr{F} = \frac{1}{2}\mathrm{Re}\left\{\nabla\left[\mathbf{p}_e^* \cdot \mathbf{E}_{\mathrm{loc}}\right]\right\}_{\mathbf{r}=\mathbf{r}_0}, \qquad (1)$$

where $\mathbf{r}_0$ is the position of the dipole center of mass. In the above, $\mathbf{E}_{\mathrm{loc}}$ represents the complex amplitude of the local field at the particle position,

$$\mathbf{E}_{\mathrm{loc}} = \mathbf{E} - \mathbf{E}^{\mathrm{self}}, \qquad (2)$$

with $\mathbf{E}$ the total electric field and $\mathbf{E}^{\mathrm{self}}$ the self-field of the dipole in free-space. It can be decomposed as:

$$\mathbf{E}_{\mathrm{loc}} = \mathbf{E}^{\mathrm{ext}} + \mathbf{E}^{\mathrm{s}}, \tag{3}$$

where $\mathbf{E}^{\mathrm{ext}}$ is the external electric field that drives the oscillations and $\mathbf{E}^{\mathrm{s}}$ is the field back-scattered by the substrate due to the dipole oscillations. Specifically, $\mathbf{E}^{\mathrm{ext}}$ is the field distribution created by the incident beam in the absence of the dipolar particle, i.e., it is the superposition of the incident wave and of the wave reflected on the substrate in the absence of the particle. On the other hand, $\mathbf{E}^{\mathrm{self}} + \mathbf{E}^{\mathrm{s}}$ is the field radiated by a particle with the dipole moment $\mathbf{p}_e$ standing above the substrate. The dipole moment $\mathbf{p}_e$ is itself a function of $\mathbf{E}_{\mathrm{loc}}$ and of the electric polarizability of the particle and its precise value must be determined self-consistently. Conversely, in principle it is possible to tailor the microstructure of the particle to obtain any desired $\mathbf{p}_e$. Here, we focus on linearly polarized dipoles, which can be readily implemented with either short metallic wires or with current loops, in the electric and magnetic cases, respectively.

From the previous discussion, the optical force can be decomposed as:

$$\mathscr{F} = \mathscr{F}^{\mathrm{ext}} + \mathscr{F}^{\mathrm{s}}, \tag{4}$$

with $\mathscr{F}^{\mathrm{ext}} = \frac{1}{2}\mathrm{Re}\left\{\nabla\left[\mathbf{p}_e^* \cdot \mathbf{E}^{\mathrm{ext}}\right]\right\}_{\mathbf{r}=\mathbf{r}_0}$ the force component due to a gradient in the external field and $\mathscr{F}^{\mathrm{s}} = \frac{1}{2}\mathrm{Re}\left\{\nabla\left[\mathbf{p}_e^* \cdot \mathbf{E}^{\mathrm{s}}\right]\right\}_{\mathbf{r}=\mathbf{r}_0}$ the force component due to the dipole oscillations.

We will consider throughout the article that the gradient of the external field vanishes along the relevant lateral test direction $\partial_i \mathbf{E}^{\mathrm{ext}} = 0$ with $\partial_i = \partial / \partial x_i$ and $i$=1 or $i$=2 ($x_1 \equiv x$ and $x_2 \equiv y$). In these conditions, the lateral force is determined simply by $\mathscr{F}^{\mathrm{s}}$. It is shown in Appendix A, that the back-scattered field can be written as $\mathbf{E}^{\mathrm{s}}(\mathbf{r}) = \mathbf{C}_{\mathrm{int}}^{\mathrm{ee}}(\mathbf{r},\mathbf{r}_0) \cdot \mathbf{p}_e / \varepsilon_0$ with $\mathbf{C}_{\mathrm{int}}^{\mathrm{ee}}$ a 3×3 dyadic expressed in terms of the reflection coefficient matrix $\mathbf{R}$ for plane wave incidence [Eq. A1b]. Therefore, in the outlined conditions the time-averaged lateral force acting on the dipole along the *i-th* direction is,

$$\mathcal{F}_i^{\mathrm{s}} = \mathrm{Re}\left\{ \frac{1}{2\varepsilon_0} \mathbf{p}_e^* \cdot \left. \frac{\partial \mathbf{C}_{\mathrm{int}}^{\mathrm{ee}}}{\partial x_i} \right|_{\mathbf{r}=\mathbf{r}_0} \cdot \mathbf{p}_e \right\}, \qquad (i=1,2). \tag{5}$$

### *B. Electric dipole standing above a uniaxial material with tilted optical axes*

To illustrate the application of the described theory, we consider that the electric dipole moment of the particle is oriented along the *z*-direction: $\mathbf{p}_e = p_e \hat{\mathbf{z}}$. The substrate is a uniaxial dielectric material with relative permittivity tensor $\boldsymbol{\varepsilon} = \varepsilon_{\alpha\alpha} \hat{\boldsymbol{\alpha}} \otimes \hat{\boldsymbol{\alpha}} + \varepsilon_{\|} \left( \mathbf{1} - \hat{\boldsymbol{\alpha}} \otimes \hat{\boldsymbol{\alpha}} \right)$, with optical axis in the *xoz* plane oriented along $\hat{\boldsymbol{\alpha}} = \sin\alpha_0 \, \hat{\mathbf{x}} + \cos\alpha_0 \, \hat{\mathbf{z}}$ (see Fig. 1b). For now, the permittivity components are taken equal to $\varepsilon_{\alpha\alpha} = -\infty$ and $\varepsilon_{\|} = 1$, which corresponds to a regime of extreme anisotropy. The tilt angle is $\alpha_0 = 45°$. To ease the visualization of the structure, in Fig. 1b the substrate is represented as a "tilted wire medium", formed by an array of tilted metallic wires [33-37]. Note that in our analysis the substrate is regarded as perfectly smooth with no intrinsic granularity. The interaction dyadic $\mathbf{C}_{\mathrm{int}}^{\mathrm{ee}}$ is given by Eq. (A1b) with the reflection matrix $\mathbf{R}$ determined as explained in Appendix B.

Figure 2a shows the far-field radiation pattern of the scattered electric field at $z = d$ plane (dashed line) for a dipole at the normalized distance $d\omega / c = 1$ from the interface. As seen, on the overall the fields are scattered more strongly towards the $x > 0$ semi-plane than towards the $x < 0$ semi-plane. The asymmetry of the radiation pattern and the conservation of linear momentum suggest that the emitter should experience a lateral recoil force directed towards the $-x$-direction. Surprisingly, by evaluating Eq. (5) using numerical methods we find that up to machine precision the lateral force $\mathcal{F}_i^{\mathrm{s}}$ is precisely zero (Fig. 2b). We numerically verified that this result is rather general and does not depend on any of the structural parameters of the problem (e.g., the direction $\alpha_0$ of the optical axis or the distance to the interface).

### *C. Reciprocity constraints*

Next, it is shown that the result of the previous subsection is rather universal and that independent of the anisotropy or chirality of the reciprocal substrate, for a linearly-polarized electric dipole the lateral force $\mathcal{F}_i^{\mathrm{s}}$ invariably vanishes. The substrate may be non-uniform along the $z$-direction (e.g., it can have finite thickness) but it should be invariant to translations along $x$ and $y$.

The proof is based on the reciprocity theorem, $\int \mathbf{E}' \cdot \mathbf{j}'' dV = \int \mathbf{E}'' \cdot \mathbf{j}' dV$, which applies to generic solutions (primed and unprimed) of Maxwell's equations that satisfy radiation boundary conditions. We take $\mathbf{E}'$ ($\mathbf{E}''$) as the field radiated by a generic electric dipole $\mathbf{j}' = -i\omega \mathbf{p}'_e \delta(\mathbf{r}-\mathbf{r}')$ ($\mathbf{j}'' = -i\omega \mathbf{p}''_e \delta(\mathbf{r}-\mathbf{r}'')$) standing above the reciprocal substrate. Since the reciprocity relation is also satisfied by the self-fields in free-space, it follows that it must also be satisfied by the back-scattered fields: $\int \mathbf{E}'^{\mathrm{s}} \cdot \mathbf{j}'' dV = \int \mathbf{E}''^{\mathrm{s}} \cdot \mathbf{j}' dV$. Using now $\mathbf{E}^{\mathrm{s}}(\mathbf{r}) = \mathbf{C}_{\mathrm{int}}^{\mathrm{ee}}(\mathbf{r},\mathbf{r}_0) \cdot \mathbf{p}_e / \varepsilon_0$ we conclude that for two generic electric dipoles $\mathbf{p}'_e$, $\mathbf{p}''_e$ positioned at $\mathbf{r}'$, $\mathbf{r}''$, respectively, one has,

$$\mathbf{p}''_e \cdot \mathbf{C}_{\mathrm{int}}^{\mathrm{ee}}(\mathbf{r}'',\mathbf{r}') \cdot \mathbf{p}'_e = \mathbf{p}'_e \cdot \mathbf{C}_{\mathrm{int}}^{\mathrm{ee}}(\mathbf{r}',\mathbf{r}'') \cdot \mathbf{p}''_e . \tag{6}$$

Suppose now that the $z$-coordinate (perpendicular to the substrate) of the two dipoles is the same: $z' = z'' \equiv z_0$. Since the substrate is invariant to translations along $x$ and $y$ directions, the interaction dyadic is necessarily of the form $\mathbf{C}_{\text{int}}^{\text{ee}}(\mathbf{r}_2, \mathbf{r}_1) = \mathbf{C}_{\text{int}}^{\text{ee}}(\mathbf{r}_2 - \mathbf{r}_1, z_0)$. Hence, taking first the derivative of both sides of Eq. (6) with respect to $x_i'$ ($i$=1,2) and then setting $\mathbf{r}_1 = \mathbf{r}_2 \equiv \mathbf{r}_0$ it follows that:

$$\mathbf{p}_e'' \cdot \left[\partial_i \mathbf{C}_{\text{int}}^{\text{ee}}\right]_{\mathbf{r}=\mathbf{r}_0} \cdot \mathbf{p}_e' = -\mathbf{p}_e' \cdot \left[\partial_i \mathbf{C}_{\text{int}}^{\text{ee}}\right]_{\mathbf{r}=\mathbf{r}_0} \cdot \mathbf{p}_e'', \qquad (7)$$

i.e., $\left[\partial_i \mathbf{C}_{\text{int}}^{\text{ee}}\right]_{\mathbf{r}=\mathbf{r}_0}$ is an anti-symmetric tensor. In particular, if one chooses $\mathbf{p}_e' = \mathbf{p}_e'' \equiv \mathbf{p}_e$ it follows that $\mathbf{p}_e \cdot \partial_i \mathbf{C}_{\text{int}} \cdot \mathbf{p}_e = 0$, and this proves that for linear polarization (when $\mathbf{p}_e$ can be taken as a real-valued vector) the lateral force $\mathscr{F}_i^{\text{s}}$ [Eq. (5)] vanishes, in agreement with the numerical results of subsection II.B. This concludes the proof that the lateral force vanishes.

It is worth noting that the reciprocity property [Eq. (6)] implies that for a dipole oriented along the $z$-direction the radiation pattern of the $z$-component of the electric field $\left|E_z^{\text{s}}\right|$ must have the parity-symmetry $(x, y, z) \to (-x, -y, z)$. Our numerical simulations (see the solid line in Fig. 2a) confirm this property. However, the reciprocity property does not enforce any particular symmetry on the *total* radiation pattern $\left|\mathbf{E}^{\text{s}}\right|^2$ (dashed line in Fig. 2a).

More generally, the reciprocity implies that the lateral force acting on a dipole with dipolar moment $\mathbf{p}_e$ differs by a minus sign from the lateral force acting on a dipole with dipole moment $\mathbf{p}_e^*$, i.e., with the opposite handiness: $\mathscr{F}_{L,\mathbf{p}_e}^{\text{s}} = -\mathscr{F}_{L,\mathbf{p}_e^*}^{\text{s}}$. This property implies, for example, that the $\mathscr{F}_i^{\mathrm{s}}$ component of the lateral force acting on a left-circularly polarized dipole is always the additive symmetric of the force acting on a right-circularly polarized dipole when the substrate is reciprocal.

A straightforward generalization of the previous analysis shows that when the particle can be modeled as a single linearly-polarized magnetic dipole the lateral force ($\mathscr{F}_i^{\mathrm{s}}$) also vanishes.

### *D. Relaxation problem*

The analysis of the previous subsections can be readily extended to the scenario of free (damped) oscillations with no external excitation. For an open system, the natural modes of oscillation of a dipole are characterized by a complex-valued frequency $\omega = \omega_0' - i\omega_0''$, with $\omega_0'' > 0$, so that the time variation is of the form $e^{-i\omega_0' t} e^{-\omega_0'' t}$. In this case, the total optical force acting on the dipole is simply

$$\mathscr{F} = \underbrace{\frac{1}{2}\mathrm{Re}\left\{\nabla\left[\mathbf{p}_e^* \cdot \mathbf{E}^{\mathrm{s}}\right]\right\}_{\mathbf{r}=\mathbf{r}_0}}_{\mathscr{F}^{\mathrm{s}}} e^{-2\omega_0'' t}. \tag{8}$$

The force $\mathscr{F}^{\mathrm{s}}$ is still determined by Eq. (5) but now $\mathbf{C}_{\mathrm{int}}^{\mathrm{ee}}$ is evaluated at the complex frequency $\omega = \omega_0' - i\omega_0''$. Note that in a relaxation problem $\omega_0' - i\omega_0''$ and $\mathbf{p}_e$ depend on the microstructure of the particle and are generally found from the solution of an eigenvalue problem.

By analytical continuation arguments, the reciprocity result (6) still holds for complex-valued values of $\omega$. Therefore, proceeding as in subsection II.C one can readily prove that when $\mathbf{p}_e$ is linearly-polarized the lateral ($x$ and $y$) components of the lateral force ($\mathscr{F}^{\mathrm{s}}$) must vanish when the substrate is reciprocal. In other words, when an excited linearly-polarized particle relaxes above a reciprocal translation-invariant substrate the optical force is always directed along the normal ($z$) direction, so that the lateral components are precisely zero.

## III. Optical Force on a Composite Linearly-Polarized Dipole

Next, we identify a new opportunity to generate a lateral force from the radiation emitted by a generic *linearly polarized* particle. Specifically, we will focus on the case wherein the particle response has both electric and magnetic components described by linearly-polarized *collinear* electric and magnetic dipoles $\mathbf{p}_e$ and $\mathbf{p}_m$, for example the particle may be formed by a short metallic wire and a small metallic loop. The directional emission of composite particles above an isotropic substrate was recently studied in Ref. [38], but the analysis was restricted to the case wherein $\mathbf{p}_e$ and $\mathbf{p}_m$ are perpendicular, i.e., to a case where the composite particle also radiates asymmetrically in free-space. In contrast, here we study the scenario wherein $\mathbf{p}_e$ and $\mathbf{p}_m$ are collinear so that the free-space radiation pattern has cylindrical (revolution) symmetry. Similar to Sect. II.A and without loss of generality, it is assumed in the following that the oscillations are driven by some external excitation.

### *A. General case*

Let us first consider the general case wherein the particle is described by arbitrary electric and magnetic dipoles $\mathbf{p}_e$ and $\mathbf{p}_m$ (our $\mathbf{p}_m$ is related to the standard magnetic dipole moment with units of $\text{A}\times\text{m}^2$ as $\mathbf{p}_m = \mu_0 \mathbf{m}$). Then, the time-averaged force can be written as [6, 39, 40]:

$$\mathscr{F} = \frac{1}{2}\mathrm{Re}\left\{\nabla\left[\mathbf{p}_g^* \cdot \mathbf{F}_{\mathrm{loc}}\right]\right\}_{\mathbf{r}=\mathbf{r}_0} - \frac{1}{12\pi}\frac{\omega^4}{c^3}\mathrm{Re}\left\{\mathbf{p}_e \times \mathbf{p}_m^*\right\}, \tag{9}$$

where $\mathbf{p}_g = \begin{pmatrix}\mathbf{p}_e & \mathbf{p}_m\end{pmatrix}^T$ and $\mathbf{F}_{\mathrm{loc}} = \begin{pmatrix}\mathbf{E}_{\mathrm{loc}} & \mathbf{H}_{\mathrm{loc}}\end{pmatrix}^T$ are six-vectors. As before, the subscript "loc" refers to the local fields. Similar to Sect. II.A, it is possible to decompose the local field as $\mathbf{F}_{\mathrm{loc}} = \mathbf{F}^{\mathrm{ext}} + \mathbf{F}^{\mathrm{s}}$ with $\mathbf{F}^{\mathrm{ext}} = \begin{pmatrix}\mathbf{E}^{\mathrm{ext}} & \mathbf{H}^{\mathrm{ext}}\end{pmatrix}^T$ the external field distribution and $\mathbf{F}^{\mathrm{s}} = \begin{pmatrix}\mathbf{E}^{\mathrm{s}} & \mathbf{H}^{\mathrm{s}}\end{pmatrix}^T$ the field back-scattered by the substrate due to the dipole radiation. Hence, for a composite dipole the total force can be written as:

$$\mathscr{F} = \mathscr{F}^{\mathrm{ext}} + \mathscr{F}^{\mathrm{s}} + \mathscr{F}^{\mathrm{self}}, \tag{10}$$

with $\mathscr{F}^{\mathrm{ext}} = \frac{1}{2}\mathrm{Re}\left\{\nabla\left[\mathbf{p}_g^* \cdot \mathbf{F}^{\mathrm{ext}}\right]\right\}_{\mathbf{r}=\mathbf{r}_0}$, $\mathscr{F}^{\mathrm{s}} = \frac{1}{2}\mathrm{Re}\left\{\nabla\left[\mathbf{p}_g^* \cdot \mathbf{F}^{\mathrm{s}}\right]\right\}_{\mathbf{r}=\mathbf{r}_0}$ and $\mathscr{F}^{\mathrm{self}} = -\frac{1}{12\pi}\frac{\omega^4}{c^3}\mathrm{Re}\left\{\mathbf{p}_e \times \mathbf{p}_m^*\right\}$. The component $\mathscr{F}^{\mathrm{self}}$ can be nonzero even when the dipole stands alone in free-space due to the asymmetric radiation of Huygens-type emitters, which does not have cylindrical symmetry [38, 41].

As in Sect. II.A, we will focus on the $\mathscr{F}^{\mathrm{s}}$ component of the force. Evidently, the field back-scattered by the substrate $\mathbf{F}^{\mathrm{s}}$ can be linked with the generalized dipole moment $\mathbf{p}_g$ through a 6×6 interaction dyadic $\mathbf{C}_g$ as $\mathbf{F}^{\mathrm{s}}(\mathbf{r}) = \mathbf{C}_g(\mathbf{r}, \mathbf{r}_0) \cdot \mathbf{p}_g$. The interaction dyadic can be decomposed as

$$\mathbf{C}_g = \begin{pmatrix} \mathbf{C}_{\mathrm{int}}^{\mathrm{ee}} / \varepsilon_0 & \mathbf{C}_{\mathrm{int}}^{\mathrm{em}} c \\ \mathbf{C}_{\mathrm{int}}^{\mathrm{me}} c & \mathbf{C}_{\mathrm{int}}^{\mathrm{mm}} / \mu_0 \end{pmatrix} \tag{11}$$

with $\mathbf{C}_{\mathrm{int}}^{\mathrm{ij}}$ (*i,j=e,m*) being $3\times3$ tensors with dimensions of $\mathrm{m}^{-3}$. The explicit formulas of $\mathbf{C}_{\mathrm{int}}^{\mathrm{ee}}$ (defined as in Sect. II.A) and $\mathbf{C}_{\mathrm{int}}^{\mathrm{mm}}$ can be found in Appendix A [Eqs. (A1b) and (A2b)]. The tensors with crossed indices can be found from the Maxwell equations (the composite dipole stands in air):

$$\mathbf{C}_{\text{int}}^{\text{me}} = \frac{1}{i(\omega/c)} \nabla \times \mathbf{C}_{\text{int}}^{\text{ee}}, \qquad \mathbf{C}_{\text{int}}^{\text{em}} = \frac{-1}{i(\omega/c)} \nabla \times \mathbf{C}_{\text{int}}^{\text{mm}}. \tag{12}$$

The force $\mathscr{F}^{\text{s}}$ is written is terms of $\mathbf{C}_g$ as follows:

$$\begin{aligned}\mathscr{F}_i^{\text{s}} &= \frac{1}{2}\text{Re}\left\{\mathbf{p}_g^* \cdot \left[\partial_i \mathbf{C}_g\right]_{\mathbf{r}=\mathbf{r}_0} \cdot \mathbf{p}_g\right\} \\ &= \frac{1}{2}\text{Re}\left\{\mathbf{p}_e^* \cdot \frac{1}{\varepsilon_0}\partial_i \mathbf{C}_{\text{int}}^{\text{ee}} \cdot \mathbf{p}_e + \mathbf{p}_m^* \cdot \frac{1}{\mu_0}\partial_i \mathbf{C}_{\text{int}}^{\text{mm}} \cdot \mathbf{p}_m + c\mathbf{p}_e^* \cdot \partial_i \mathbf{C}_{\text{int}}^{\text{em}} \cdot \mathbf{p}_m + c\mathbf{p}_m^* \cdot \partial_i \mathbf{C}_{\text{int}}^{\text{me}} \cdot \mathbf{p}_e\right\}_{\mathbf{r}=\mathbf{r}_0}\end{aligned} \tag{13}$$

When the external driving field has zero gradient along the lateral direction of interest, the lateral force is completely determined by $\mathscr{F}^{\text{s}} + \mathscr{F}^{\text{self}}$.

### *B. Reciprocity constraints*

The general form of the reciprocity theorem when the electromagnetic sources have both electric and magnetic components is [42]

$$\int \mathbf{j}'' \cdot \boldsymbol{\sigma}_z \cdot \mathbf{F}' dV = \int \mathbf{j}' \cdot \boldsymbol{\sigma}_z \cdot \mathbf{F}'' dV, \tag{14}$$

with $\mathbf{F} = \begin{pmatrix}\mathbf{E} & \mathbf{H}\end{pmatrix}^T$ the six-vector that determines the electromagnetic field, $\mathbf{j} = \begin{pmatrix}\mathbf{j}_e & \mathbf{j}_m\end{pmatrix}^T$ the six-vector formed by the electric and magnetic current densities and $\boldsymbol{\sigma}_z = \begin{pmatrix}\mathbf{1}_{3\times3} & 0 \\ 0 & -\mathbf{1}_{3\times3}\end{pmatrix}$ is a matrix with dimension 6. Thus, considering the distributions of fields $(\mathbf{F}', \mathbf{F}'')$ created by the current distributions $\mathbf{j}' = -i\omega \mathbf{p}_g' \delta(\mathbf{r} - \mathbf{r}')$ and $\mathbf{j}'' = -i\omega \mathbf{p}_g'' \delta(\mathbf{r} - \mathbf{r}'')$ and proceeding as in Sect. II.C it is possible to prove that:

$$\mathbf{p}_g'' \cdot \boldsymbol{\sigma}_z \cdot \mathbf{C}_g(\mathbf{r}'', \mathbf{r}') \cdot \mathbf{p}_g' = \mathbf{p}_g' \cdot \boldsymbol{\sigma}_z \cdot \mathbf{C}_g(\mathbf{r}', \mathbf{r}'') \cdot \mathbf{p}_g''. \tag{15}$$

Furthermore, for a structure invariant to translations along $x$ and $y$ one may find that for arbitrary $\mathbf{p}'_g$, $\mathbf{p}''_g$ the following identity holds,

$$\mathbf{p}''_g \cdot \boldsymbol{\sigma}_z \cdot \left[\partial_i \mathbf{C}_g\right]_{\mathbf{r}=\mathbf{r}_0} \cdot \mathbf{p}'_g = -\mathbf{p}'_g \cdot \boldsymbol{\sigma}_z \cdot \left[\partial_i \mathbf{C}_g\right]_{\mathbf{r}=\mathbf{r}_0} \cdot \mathbf{p}''_g, \qquad (\underline{i}=x,y) \qquad (16)$$

Picking $\mathbf{p}'_g = \mathbf{p}''_g = \begin{pmatrix}\mathbf{p}_e & 0\end{pmatrix}^T$ we see that $\mathbf{p}_e \cdot \frac{1}{\varepsilon_0}\partial_i \mathbf{C}_{\text{int}}^{\text{ee}} \cdot \mathbf{p}_e = 0$ and picking $\mathbf{p}'_g = \mathbf{p}''_g = \begin{pmatrix}0 & \mathbf{p}_m\end{pmatrix}^T$ we get $\mathbf{p}_m \cdot \frac{1}{\mu_0}\partial_i \mathbf{C}_{\text{int}}^{\text{mm}} \cdot \mathbf{p}_m = 0$. Furthermore, choosing $\mathbf{p}'_g = \begin{pmatrix}\mathbf{p}_e & 0\end{pmatrix}^T$ and $\mathbf{p}''_g = \begin{pmatrix}0 & \mathbf{p}_m\end{pmatrix}^T$ one finds that $\mathbf{p}_m \cdot \partial_i \mathbf{C}_{\text{int}}^{\text{me}} \cdot \mathbf{p}_e = \mathbf{p}_e \cdot \partial_i \mathbf{C}_{\text{int}}^{\text{em}} \cdot \mathbf{p}_m$.

Let us now suppose that both the electric and magnetic dipoles are *linearly-polarized* so that $\mathbf{p}_e = |p_e| e^{i\phi_e}\hat{\mathbf{u}}_e$ and $\mathbf{p}_m = |p_m| e^{i\phi_m}\hat{\mathbf{u}}_m$ with the unit vectors $\hat{\mathbf{u}}_e, \hat{\mathbf{u}}_m$ real-valued. Using the reciprocity constraints derived in the previous paragraph in Eq. (13), it follows that, the lateral force $\mathscr{F}_i^{\text{s}}$ reduces to:

$$\mathscr{F}_i^{\text{s}} = \cos(\Delta) c |p_e||p_m| \operatorname{Re}\left\{\hat{\mathbf{u}}_m \cdot \partial_i \mathbf{C}_{\text{int}}^{\text{me}} \cdot \hat{\mathbf{u}}_e\right\}_{\mathbf{r}=\mathbf{r}_0} \qquad \text{(lin. pol. dipoles)} \qquad (17)$$

with $\Delta = \phi_e - \phi_m$ the phase difference between the two dipole moments. Therefore, for a composite dipole the lateral force $\mathscr{F}_i^{\text{s}}$ may be nonzero, even when the dipoles are linearly polarized. The lateral force has maximal strength when the two dipoles oscillate in phase ($\Delta = 0$) or in opposition of phase ($\Delta = 180^{\text{o}}$), and vanishes when they are in quadrature ($\Delta = \pm 90^{\text{o}}$).

## *C. Collinear linearly-polarized dipoles*

The case of collinear dipoles ($\hat{\mathbf{u}}_e = \hat{\mathbf{u}}_m$) is particularly interesting because the corresponding self-force $\mathscr{F}^{\text{self}}$ vanishes. Thus, when the gradient of the external field along the lateral direction of interest is zero, the lateral force is fully determined by $\mathscr{F}^{s}$ [Eq. (17)]. Using $\mathbf{C}_{\rm int}^{\rm me} = \frac{1}{i(\omega/c)} \nabla \times \mathbf{C}_{\rm int}^{\rm ee}$ and Eq. (A1b) it is possible to obtain an explicit formula for the lateral force, but it is a bit too cumbersome to show here. When both dipoles are perpendicular to the interface ($\hat{\mathbf{u}}_e = \hat{\mathbf{u}}_m = \hat{\mathbf{z}}$) and the composite particle is positioned at a distance $d$ from the substrate we get:

$$\frac{\mathscr{F}_L^s}{\mathscr{F}_0} = \cos(\Delta) \mathrm{Re} \left\{ \frac{d^4}{(2\pi)^2} \iint dk_x dk_y \frac{1}{2} \frac{-\mathbf{k}_t}{(\omega/c)} e^{-\gamma_0 2d} (\hat{\mathbf{z}} \times \mathbf{k}_t) \cdot \mathbf{R}(\omega, k_x, k_y) \cdot \mathbf{k}_t \right\} \quad (18)$$

with $\mathscr{F}_0 = c|p_e||p_m|/d^4$ a normalizing factor with units of force (N), $\mathbf{R}(\omega, k_x, k_y)$ the reflection matrix for plane wave incidence, and $\mathbf{k}_t = k_x \hat{\mathbf{x}} + k_y \hat{\mathbf{y}}$ the transverse wave vector (see Appendix A).

### *D. Uniaxial dielectric substrate*

To illustrate the possibilities, next we consider a composite particle formed by the superposition of two collinear (linearly-polarized) electric and magnetic dipoles positioned at a distance $d$ above the same uniaxial dielectric substrate as in Sect. II.B. For simplicity, the transverse permittivity of the uniaxial dielectric is taken equal to $\varepsilon_{||} = 1$ in all the simulations.

Even though the reciprocity constraints do not forbid a lateral force for a superposition of electric and magnetic dipoles, surprisingly, it turns out that the force component along the *x*–direction vanishes. Even more puzzling, we find that the lateral force can be non-zero along the *y*-direction. Figures 3a and 3b show how the *y*-component of the lateral optical force $\mathscr{F}_y^{s}$ varies with the normalized distance $d\omega/c$ and with the tilt angle of the optical axis $\alpha_0$. As seen in Fig. 3a, the sign of the lateral force can be tuned by varying the height of the particle with respect to the substrate, so that the force can be either positive or negative. The lateral force depends strongly on the value of $\varepsilon_{\alpha\alpha}$, and most notably its sign changes when $\varepsilon_{\alpha\alpha}$ crosses the point $\varepsilon_{\alpha\alpha} = -1$. Furthermore, the lateral force depends on the tilt angle of the substrate optical axis (Fig. 3b), and vanishes when the optical axis is either normal or parallel to the interface. The lateral force sign is locked to the tilt angle sign.

Figure 3c illustrates how the lateral force varies with the orientation of the two collinear dipoles $\alpha_e = \alpha_m$ for the normalized distance $d\omega / c = 1$ and for $\alpha_0 = 45^{\text{o}}$. Curiously, when $\varepsilon_{\|} = 1$ as considered here, the lateral force vanishes when the dipole moment is parallel to the optical axis direction $\alpha_e = \alpha_m = \alpha_0$. In agreement with Eq. (17), Fig. 3d shows that the lateral force depends on the phase difference, $\Delta = \phi_e - \phi_m$, between the two dipole moments, and has the largest value when the dipole moments are either in phase or in opposition of phase.

Hereafter, we consider the case where the two dipole moments are vertical ($\alpha_e = \alpha_m = 0$) and oscillate in phase ($\Delta = 0$). Furthermore, it is assumed that the normalized distance satisfies $d\omega / c = 1$, and that the substrate has $\alpha_0 = 45^{\text{o}}$ and $\varepsilon_{\alpha\alpha} = -\infty$ (extreme anisotropy), corresponding to a positive lateral force (Fig. 3) along the +*y*-direction.

Figure 4 shows a polar plot of the back-scattered electric field pattern $\left|\mathbf{E}^{s}\right|^2$ in the half-space $z > 0$ at $5\lambda_0$ radial distance from the composite particle (red line). For the sake of completeness, we also show the back-scattered field patterns when the particle is a simple electric dipole (green line) or a simple magnetic dipole (blue line).

The composite particle has a radiation pattern that is quasi-symmetric in the *y=0* plane (Fig. 4b). On the other hand, in the $x=0$ plane the radiation from the particle is strongly scattered to the negative *y* direction (Fig. 4c) as expected from the lateral force being positive ($\mathscr{F}_y^{s} > 0$). This property is further highlighted in Fig. 4a, which shows that in the *xoy* plane the fields are scattered almost equally to the right and left but mainly to the negative *y* semi-plane. Figure 5 shows the near-field density plots of $\left|\mathbf{E}^{s}\right|^2$ in different cut-planes. In agreement with Fig. 4, the density plots are strongly asymmetric in the *yoz* plane.

To unveil the reason why the force is directed along *y*, it is convenient to visualize the uniaxial material as a tilted wire medium (Fig. 4d) [33-36]. In these conditions, when the two collinear (vertical) dipoles are placed above the substrate, they induce a current along the metallic wires. The combined effect of the dipoles and of the current induced along the wires is arguably analogous to a Huygens radiator with the effective electric and magnetic dipoles in the *xoz* plane (Fig. 4d). This elementary analogy explains why the radiation pattern may be strongly asymmetric along the *y*-direction when $\alpha_e = \alpha_m \neq \alpha_0$, despite the structural symmetry of the substrate. Furthermore, Fig. 4d elucidates why the lateral force vanishes when $\alpha_e = \alpha_m = \alpha_0$ (Fig. 3c), because in this scenario the effective electric and magnetic dipoles are evidently parallel. A different explanation why for $\alpha_e = \alpha_m \neq \alpha_0$ the lateral force can be nontrivial is that the dipole-emitter is not invariant under the parity transformation $(x,y,z) \to (x,-y,z)$ because this transformation flips the magnetic dipole but not the electric dipole. This breach of symmetry explains the asymmetric radiation patterns along the $y$-direction.

## IV. Conclusion

In summary, we theoretically studied the lateral (recoil) forces acting on generic dipole type particles placed nearby a reciprocal translation-invariant substrate. It was shown that counter-intuitively the scattering from electric-dipoles above a smooth reciprocal planar substrate does not lead to recoil-type forces, even if the substrate is chiral or has tilted optical axes. A nontrivial lateral force can only be generated when the external field excitation has a gradient along a direction parallel to the interface.

We proposed a novel mechanism to create lateral forces using a composite dipolar particle formed by collinear linearly-polarized electric and magnetic dipoles, e.g., a composite particle formed by short metallic wire and a metal loop. We discussed in detail the physical mechanisms that enable the emergence of the anomalous recoil force when the composite particle stands above a uniaxial material with tilted optical axes. Our results may enable exotic optical manipulations and may find applications in particle sorting and delivery.

**Acknowledgement:** This work is supported in part by the IET under the Harvey Engineering Research Prize, by Fundação para a Ciência e a Tecnologia grant number PTDC/EEI-TEL/4543/2014 and by Instituto de Telecomunicações under project UID/EEA/50008/2017.

## Appendix A: The back-scattered field

In this Appendix, we obtain the fields back-scattered by a material substrate for a dipole-type excitation. The electric and magnetic dipoles cases are analyzed separately.

The fields radiated by a composite dipole (with electric and magnetic components) can be simply obtained using the superposition principle.

**Electric dipole**

Next, we characterize the fields radiated by an electric dipole that oscillates with frequency $\omega$ in air (region $z>0$) in the vicinity of a planar material structure (region $z<0$) (Fig. 1a). The electromagnetic fields in the $z>0$ region are the superposition of the field radiated by the dipole in free-space ($\mathbf{E}^{\text{self}}$) and the field back-scattered by the material slab ($\mathbf{E}^{\text{s}}$), $\mathbf{E}=\mathbf{E}^{\text{self}}+\mathbf{E}^{\text{s}}$. The self-electric field is given by $\mathbf{E}^{\text{self}}=\nabla\times\nabla\times\left(\frac{\mathbf{p}_e}{\varepsilon_0}\Phi_0\right)$ for $\mathbf{r}\neq\mathbf{r}_0$, where $\mathbf{r}_0=\left(x_0,y_0,z_0\right)$ determines the coordinates of the electric dipole, $\mathbf{p}_e$ is the electric dipole moment, $\Phi_0=\frac{e^{ik_0|\mathbf{r}-\mathbf{r}'|}}{4\pi|\mathbf{r}-\mathbf{r}'|}$ is the Hertz potential, and $k_0=\omega/c$ is the free-space wave number.

Let us consider the problem of plane wave incidence on the material slab with the incident wave propagating in the air region ($z>0$). Let $\mathbf{k}_t=k_x\hat{\mathbf{x}}+k_y\hat{\mathbf{y}}$ be the transverse (parallel to the interface) wave vector of the incident plane wave that determines the variation of the fields ($e^{i\mathbf{k}_t\cdot\mathbf{r}}$) along the $x$ and $y$ directions. We introduce $\mathbf{R}=\mathbf{R}\left(\omega,k_x,k_y\right)$ as the (2×2) reflection matrix such that the transverse electric field ($\mathbf{E}_t^{\text{ref}}=E_x^{\text{ref}}\hat{\mathbf{x}}+E_y^{\text{ref}}\hat{\mathbf{y}}$) associated with the plane wave reflected on the material slab is related to the transverse electric field associated with the incident plane wave ($\mathbf{E}_t^{\text{inc}}=E_x^{\text{inc}}\hat{\mathbf{x}}+E_y^{\text{inc}}\hat{\mathbf{y}}$) as $\mathbf{E}_t^{\text{ref}}=\mathbf{R}\cdot\mathbf{E}_t^{\text{inc}}$.

Following Refs. [28, 43], the electric field back-scattered by the material slab, $\mathbf{E}^{\mathrm{s}}$, can be written in the air region in terms of $\mathbf{R}$ through a Sommerfeld-type integral:

$$\mathbf{E}^{\mathrm{s}}(\mathbf{r}) = \mathbf{C}_{\mathrm{int}}^{\mathrm{ee}} \cdot \mathbf{p}_e / \varepsilon_0 \tag{A1a}$$

$$\mathbf{C}_{\mathrm{int}}^{\mathrm{ee}}(\mathbf{r},\mathbf{r}_0) = \frac{1}{(2\pi)^2} \iint dk_x dk_y \frac{1}{2\gamma_0} e^{-\gamma_0(z+z_0)} e^{ik_x(x-x_0)} e^{ik_y(y-y_0)} \times \left(\mathbf{1}_t + \hat{\mathbf{z}} \otimes \frac{i\mathbf{k}_t}{\gamma_0}\right) \cdot \mathbf{R} \cdot \left(i\gamma_0 \mathbf{k}_t \otimes \hat{\mathbf{z}} + k_0^2 \mathbf{1}_t - \mathbf{k}_t \otimes \mathbf{k}_t\right) \tag{A1b}$$

Here, $\gamma_0 = -ik_{z0}$ with $k_{z0} = \sqrt{\omega^2 / c^2 - k_x^2 - k_y^2}$ is the propagation constant in the air region and $\mathbf{1}_t = \hat{\mathbf{x}} \otimes \hat{\mathbf{x}} + \hat{\mathbf{y}} \otimes \hat{\mathbf{y}}$. Equation (A1) is completely general and applies to any planar substrate that is invariant to translations along the *x* and *y* directions. The magnetic field radiated by the electric dipole in the air region can be written as $\mathbf{H} = \mathbf{H}^{\mathrm{self}} + \mathbf{H}^{\mathrm{s}}$ with $\mathbf{H}^i = \frac{1}{i\omega\mu_0} \nabla \times \mathbf{E}^i$ (*i*=s, self).

**Magnetic dipole**

The fields radiated by a magnetic dipole (with dipole moment $\mathbf{p}_m$) can be easily found using "duality" arguments. Specifically, in the air region it is possible to write $\mathbf{H} = \mathbf{H}^{\mathrm{self}} + \mathbf{H}^{\mathrm{s}}$ with $\mathbf{H}^{\mathrm{self}} = \nabla \times \nabla \times \left(\frac{\mathbf{p}_m}{\mu_0} \Phi_0\right)$ for $\mathbf{r} \neq \mathbf{r}_0$ and $\mathbf{H}^{\mathrm{s}}$ given by

$$\mathbf{H}^{\mathrm{s}}(\mathbf{r}) = \mathbf{C}_{\mathrm{int}}^{\mathrm{mm}} \cdot \mathbf{p}_m / \mu_0 \tag{A2a}$$

$$\mathbf{C}_{\mathrm{int}}^{\mathrm{mm}}(\mathbf{r},\mathbf{r}_0) = \frac{1}{(2\pi)^2} \iint dk_x dk_y \frac{1}{2\gamma_0} e^{-\gamma_0(z+z_0)} e^{ik_x(x-x_0)} e^{ik_y(y-y_0)} \times \left(\mathbf{1}_t + \hat{\mathbf{z}} \otimes \frac{i\mathbf{k}_t}{\gamma_0}\right) \cdot \mathbf{R}^H \cdot \left(i\gamma_0 \mathbf{k}_t \otimes \hat{\mathbf{z}} + k_0^2 \mathbf{1}_t - \mathbf{k}_t \otimes \mathbf{k}_t\right) \tag{A2b}$$

Note that the tensor $\mathbf{C}_{\text{int}}^{\text{mm}}$ is defined in the same manner as the tensor $\mathbf{C}_{\text{int}}^{\text{ee}}$, except that the reflection coefficient matrix is different: $\mathbf{R} \to \mathbf{R}^H$. The reflection coefficient matrix $\mathbf{R}^H$ is such that for plane wave incidence the transverse components of the incident and reflected magnetic fields are linked as $\mathbf{H}_t^{\text{ref}} = \mathbf{R}^H \cdot \mathbf{H}_t^{\text{inc}}$. Using the Maxwell equations, it is straightforward to check that for a plane wave propagating in free-space the transverse electric and magnetic fields are related by:

$$\mathbf{E}_t^{\pm} = -\frac{\pm 1}{\omega \varepsilon_0 k_{z0}} \begin{pmatrix} -k_x k_y & -\left(k_y^2 + k_{z0}^2\right) \\ k_x^2 + k_{z0}^2 & k_x k_y \end{pmatrix} \cdot \mathbf{H}_t^{\pm}. \tag{A3}$$

The $\pm$ sign determines if the wave propagates towards $+z$ or $-z$ semi-space. From this result it is straightforward to show that the reflection coefficient matrix for the magnetic field can be written as,

$$\mathbf{R}^H\left(\omega, k_x, k_y\right) = -\begin{pmatrix} -k_x k_y & -\left(k_y^2 + k_{z0}^2\right) \\ k_x^2 + k_{z0}^2 & k_x k_y \end{pmatrix}^{-1} \cdot \mathbf{R} \cdot \begin{pmatrix} -k_x k_y & -\left(k_y^2 + k_{z0}^2\right) \\ k_x^2 + k_{z0}^2 & k_x k_y \end{pmatrix}. \tag{A4}$$

In the air region, the electric field radiated by the magnetic dipole is given by $\mathbf{E} = \mathbf{E}^{\text{self}} + \mathbf{E}^{\text{s}}$ with $\mathbf{E}^i = -\frac{1}{i\omega\varepsilon_0} \nabla \times \mathbf{H}^i$ ($i$=s, self).

## Appendix B: Reflection matrix for a uniaxial substrate with a tilted optical axis

Next, we derive the reflection matrix $\mathbf{R}$ (defined as in Appendix A) for the scenario wherein the region $z < 0$ is a nonmagnetic uniaxial dielectric with optical axis directed along the unit vector $\hat{\boldsymbol{\alpha}}$. The relative permittivity of the uniaxial dielectric is of the form:

$$\boldsymbol{\varepsilon} = \varepsilon_{\parallel}\left(\mathbf{1} - \hat{\boldsymbol{\alpha}} \otimes \hat{\boldsymbol{\alpha}}\right) + \varepsilon_{\alpha\alpha} \hat{\boldsymbol{\alpha}} \otimes \hat{\boldsymbol{\alpha}}. \tag{B1}$$

**Reflection matrix**

Here, we obtain a general formula for $\mathbf{R}$ in terms of wave admittance matrices. We follow closely the ideas of Ref. [44 , Ap. A], and define the transverse fields,

$$\mathbf{E}_t = \begin{pmatrix} E_x \\ E_y \end{pmatrix} \quad \text{and} \quad \mathbf{J}\cdot\mathbf{H}_t = \begin{pmatrix} 0 & 1 \\ -1 & 0 \end{pmatrix}\begin{pmatrix} H_x \\ H_y \end{pmatrix} = \begin{pmatrix} H_y \\ -H_x \end{pmatrix}. \tag{B2}$$

For some generic bulk material, it is useful to introduce two admittance matrices, $\mathbf{Y}^{\pm}$, such that for plane waves propagating along the $+z$ and $-z$ directions with transverse wave vector $\mathbf{k}_t$, the corresponding transverse fields are related by,

$$\mathbf{J}\cdot\mathbf{H}_t^{+} = \mathbf{Y}^{+}\cdot\mathbf{E}_t^{+}, \qquad \mathbf{J}\cdot\mathbf{H}_t^{-} = -\mathbf{Y}^{-}\cdot\mathbf{E}_t^{-}. \tag{B3}$$

The matrices $\mathbf{Y}^{\pm}$ depend on the material, on the frequency $\omega$, and on the transverse wave vector $\mathbf{k}_t$. For an isotropic dielectric with permittivity $\varepsilon$, one easily finds that,

$$\mathbf{Y}^{+}\eta_0 = \mathbf{Y}^{-}\eta_0 = \frac{1}{k_0 k_z^{+}}\begin{bmatrix} k_0^2\varepsilon - k_y^2 & k_x k_y \\ k_x k_y & k_0^2\varepsilon - k_x^2 \end{bmatrix} \quad \text{(isotropic dielectric)}. \tag{B4}$$

where $k_z^{+} = \sqrt{k_0^2\varepsilon - k_x^2 - k_y^2}$, $k_0 = \omega / c$ is the free-space wave number, and $\eta_0$ is the vacuum impedance.

The reflection coefficient matrix for an interface between air ($z > 0$) and a uniaxial material ($z < 0$) can be found by imposing the continuity of $\mathbf{E}_t$ and $\mathbf{J}\cdot\mathbf{H}_t$ at the interface. This procedure yields [44]:

$$\mathbf{R}(\omega, k_x, k_y) = \left(\mathbf{Y}_d + \mathbf{Y}_u^{-}\right)^{-1}\cdot\left(\mathbf{Y}_d - \mathbf{Y}_u^{-}\right). \tag{B5}$$

where $\mathbf{Y}_d = \mathbf{Y}_d^{\pm}$ is the admittance matrix of the air region [Eq. (B4) with $\varepsilon = 1$], and $\mathbf{Y}_u^-$ is the admittance matrix of the uniaxial medium. Note that in general the matrices $\mathbf{Y}_d$ and $\mathbf{Y}_u^-$ do not commute. In the next subsections, we obtain an explicit formula for $\mathbf{Y}_u^-$.

**Admittance matrices**

The admittance matrices $\mathbf{Y}^{\pm}\left(\omega, k_x, k_y\right)$ can be constructed from the plane wave modes of the bulk medium. Specifically, let $\left(\mathbf{E}_i^{\pm} \quad \mathbf{H}_i^{\pm}\right)$ ($i$=1,2) be the fields associated with the plane waves with a wave vector of the form $\mathbf{k}_i = \left(k_x, k_y, k_{z,i}^{\pm}\right)$ and $\frac{1}{2}\mathrm{Re}\left\{\mathbf{E}_i^{\pm} \times \mathbf{H}_i^{\pm,*}\right\} \cdot \left(\pm\hat{\mathbf{z}}\right) > 0$ so that the energy flows towards the ±$z$-direction depending on the superscript. Then, from Eq. (B3) we can write:

$$\left[\mathbf{J}\cdot\mathbf{H}_{t,1}^{\pm} \quad \mathbf{J}\cdot\mathbf{H}_{t,2}^{\pm}\right] = \pm\mathbf{Y}^{\pm}\cdot\left[\mathbf{E}_{t,1}^{\pm} \quad \mathbf{E}_{t,2}^{\pm}\right]. \tag{B6}$$

where the subscript "t" indicates that only the transverse components are considered. Thereby, the admittance matrices are given by:

$$\mathbf{Y}^{\pm} = \pm\left[\mathbf{J}\cdot\mathbf{H}_{t,1}^{\pm} \quad \mathbf{J}\cdot\mathbf{H}_{t,2}^{\pm}\right]\cdot\left[\mathbf{E}_{t,1}^{\pm} \quad \mathbf{E}_{t,2}^{\pm}\right]^{-1} \tag{B7}$$

For nonmagnetic structures, the plane wave fields satisfy $\mathbf{H} = \frac{1}{\omega\mu_0}\mathbf{k}\times\mathbf{E}$. Using this property it is possible to show that,

$$\mathbf{Y}^{+}\eta_0 = \frac{1}{k_0}\frac{1}{\left(\mathbf{E}_1^{+}\times\mathbf{E}_2^{+}\right)\cdot\hat{\mathbf{z}}}\begin{bmatrix} \mathbf{E}_1^{+}\cdot\left[\hat{\mathbf{y}}\left(\mathbf{k}_2^{+}\times\hat{\mathbf{y}}\right)-\left(\mathbf{k}_1^{+}\times\hat{\mathbf{y}}\right)\hat{\mathbf{y}}\right]\cdot\mathbf{E}_2^{+} & \mathbf{E}_1^{+}\cdot\left[\left(\mathbf{k}_1^{+}\times\hat{\mathbf{y}}\right)\hat{\mathbf{x}}-\hat{\mathbf{x}}\left(\mathbf{k}_2^{+}\times\hat{\mathbf{y}}\right)\right]\cdot\mathbf{E}_2^{+} \\ \mathbf{E}_1^{+}\cdot\left[\left(\mathbf{k}_1^{+}\times\hat{\mathbf{x}}\right)\hat{\mathbf{y}}-\hat{\mathbf{y}}\left(\mathbf{k}_2^{+}\times\hat{\mathbf{x}}\right)\right]\cdot\mathbf{E}_2^{+} & \mathbf{E}_1^{+}\cdot\left[\hat{\mathbf{x}}\left(\mathbf{k}_2^{+}\times\hat{\mathbf{x}}\right)-\left(\mathbf{k}_1^{+}\times\hat{\mathbf{x}}\right)\hat{\mathbf{x}}\right]\cdot\mathbf{E}_2^{+} \end{bmatrix}$$

(B8a)

A similar formula is obtained for $\mathbf{Y}^-$ with a minus sign in front of the right-hand side expression and with $\mathbf{E}_i^+$ (the waves that propagate along +*z*) replaced by $\mathbf{E}_i^-$ (the waves that propagate along –*z*):

$$\mathbf{Y}^-\eta_0 = -\frac{1}{k_0}\frac{1}{\left(\mathbf{E}_1^- \times \mathbf{E}_2^-\right)\cdot\hat{\mathbf{z}}}\begin{bmatrix} \mathbf{E}_1^- \cdot\left[\hat{\mathbf{y}}\left(\mathbf{k}_2^- \times \hat{\mathbf{y}}\right) - \left(\mathbf{k}_1^- \times \hat{\mathbf{y}}\right)\hat{\mathbf{y}}\right]\cdot \mathbf{E}_2^- & \mathbf{E}_1^- \cdot\left[\left(\mathbf{k}_1^- \times \hat{\mathbf{y}}\right)\hat{\mathbf{x}} - \hat{\mathbf{x}}\left(\mathbf{k}_2^- \times \hat{\mathbf{y}}\right)\right]\cdot \mathbf{E}_2^- \\ \mathbf{E}_1^- \cdot\left[\left(\mathbf{k}_1^- \times \hat{\mathbf{x}}\right)\hat{\mathbf{y}} - \hat{\mathbf{y}}\left(\mathbf{k}_2^- \times \hat{\mathbf{x}}\right)\right]\cdot \mathbf{E}_2^- & \mathbf{E}_1^- \cdot\left[\hat{\mathbf{x}}\left(\mathbf{k}_2^- \times \hat{\mathbf{x}}\right) - \left(\mathbf{k}_1^- \times \hat{\mathbf{x}}\right)\hat{\mathbf{x}}\right]\cdot \mathbf{E}_2^- \end{bmatrix}$$

(B8b)

When the medium is either reciprocal or invariant under a parity transformation ($\mathbf{r} \rightarrow -\mathbf{r}$), the admittance matrices have the symmetries discussed in Ref. [44].

**Uniaxial dielectric**

Let us now focus on uniaxial dielectrics with relative permittivity tensor as in Eq. (B1). In this case, the relevant plane waves are the usual ordinary ($\mathbf{E}_1$) and extraordinary ($\mathbf{E}_2$) waves, with electric field of the form (apart from an arbitrary multiplication factor):

$$\mathbf{E}_1 = \mathbf{k}_1 \times \hat{\boldsymbol{\alpha}}, \qquad \mathbf{k}_1 \equiv \mathbf{k}_O, \tag{B9a}$$

$$\mathbf{E}_2 = \boldsymbol{\varepsilon}^{-1}\cdot\left[\mathbf{k}_2 \times \left(\mathbf{k}_2 \times \hat{\boldsymbol{\alpha}}\right)\right] = \frac{\left(\mathbf{k}_2 \cdot \hat{\boldsymbol{\alpha}}\right)}{\varepsilon_{||}}\mathbf{k}_2 - \hat{\boldsymbol{\alpha}}\frac{\omega^2}{c^2}, \qquad \mathbf{k}_2 \equiv \mathbf{k}_E. \tag{B9b}$$

The ordinary wave satisfies the dispersion equation:

$$\mathbf{k}_O \cdot \mathbf{k}_O = \left(\frac{\omega}{c}\right)^2 \varepsilon_{||}, \tag{B10a}$$

while the extraordinary wave satisfies,

$$\frac{1}{\varepsilon_{||}}\left(\mathbf{k}_E \cdot \hat{\boldsymbol{\alpha}}\right)^2 + \frac{1}{\varepsilon_{\alpha\alpha}}\left[\left(\mathbf{k}_E \cdot \mathbf{k}_E\right) - \left(\mathbf{k}_E \cdot \hat{\boldsymbol{\alpha}}\right)^2\right] = \left(\frac{\omega}{c}\right)^2. \tag{B10b}$$

The second identity in Eq. (B9b) is obtained with the help of Eq. (B10b). The wave vectors of the ordinary and extraordinary waves are the form:

$$\mathbf{k}_O = k_x\hat{\mathbf{x}} + k_y\hat{\mathbf{y}} + k_{O,z}\hat{\mathbf{z}}, \qquad \mathbf{k}_E = k_x\hat{\mathbf{x}} + k_y\hat{\mathbf{y}} + k_{E,z}\hat{\mathbf{z}} \tag{B11}$$

with the $z$-component of the wave vectors determined by Eqs. (B10a) and (B10b), respectively. Straightforward calculations show that:

$$k_{O,z} = \pm\sqrt{\left(\frac{\omega}{c}\right)^2 \varepsilon_{\parallel} - k_x^2 - k_y^2} \tag{B12a}$$

$$k_{E,z} = \frac{1}{\varepsilon_{\parallel} + (\hat{\mathbf{z}}\cdot\hat{\boldsymbol{\alpha}})^2(\varepsilon_{\alpha\alpha} - \varepsilon_{\parallel})}\Big[-(\varepsilon_{\alpha\alpha} - \varepsilon_{\parallel})(\mathbf{k}_t\cdot\hat{\boldsymbol{\alpha}})(\hat{\mathbf{z}}\cdot\hat{\boldsymbol{\alpha}})$$
$$\pm\left[\left((\varepsilon_{\alpha\alpha} - \varepsilon_{\parallel})(\mathbf{k}_t\cdot\hat{\boldsymbol{\alpha}})(\hat{\mathbf{z}}\cdot\hat{\boldsymbol{\alpha}})\right)^2 + \left(\frac{\omega^2}{c^2}\varepsilon_{\alpha\alpha}\varepsilon_{\parallel} - \varepsilon_{\parallel}k_t^2 + (\mathbf{k}_t\cdot\hat{\boldsymbol{\alpha}})^2(\varepsilon_{\parallel} - \varepsilon_{\alpha\alpha})\right)\left(\varepsilon_{\parallel} + (\hat{\mathbf{z}}\cdot\hat{\boldsymbol{\alpha}})^2(\varepsilon_{\alpha\alpha} - \varepsilon_{\parallel})\right)\right]^{1/2}\Big] \tag{B12b}$$

with $\mathbf{k}_t = k_x\hat{\mathbf{x}} + k_y\hat{\mathbf{y}}$ the transverse (parallel to the interface with air) wave vector. The ± sign is chosen depending on the desired propagation direction ($\pm z$). In particular, the matrix $\mathbf{Y}_u^-$ is calculated using Eq. (B8b) with $\mathbf{E}_i^-$ defined as in Eq. (B9) with the wave vectors determined by Eq. (B12) with the "$-$" sign.

# Figures

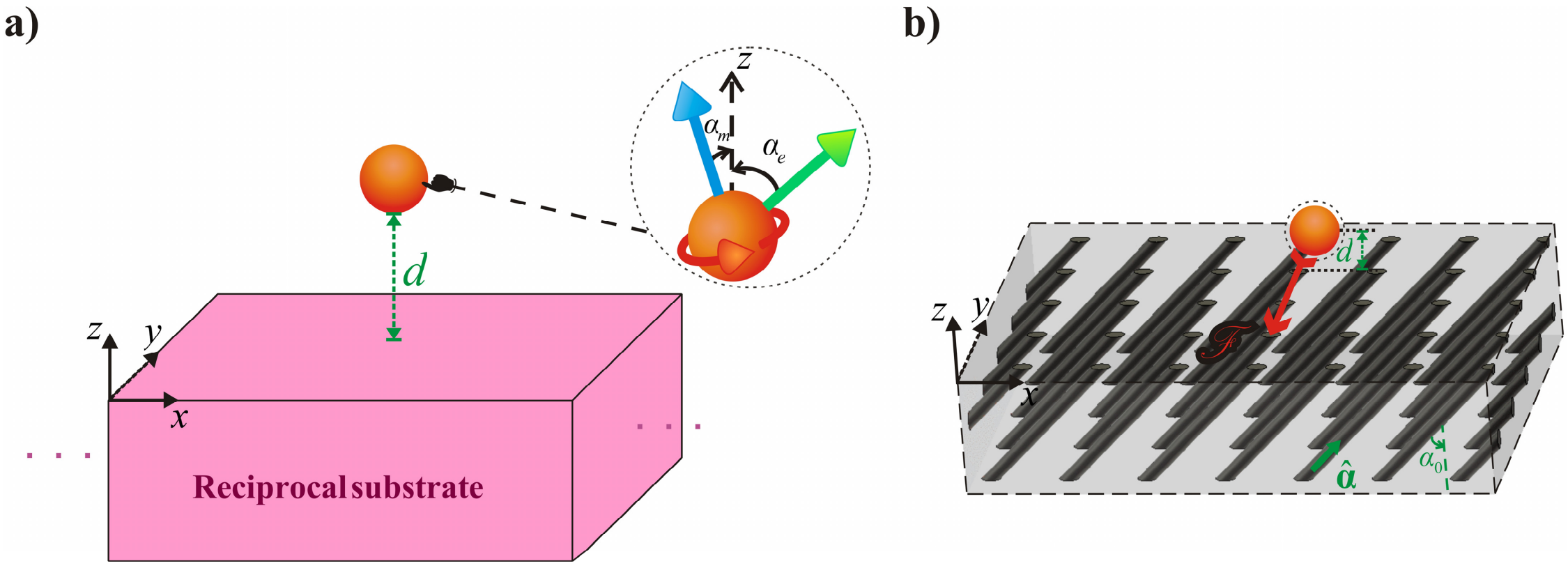

**Fig. 1**. **(a)** Geometry of the problem: A particle stands in the air region at a distance $d$ from a reciprocal material (half-space $z < 0$ ). The particle is either formed by a single electric dipole or by a superposition of linearly-polarized electric and magnetic dipoles. **(b)** Illustration of the uniaxial material substrate with tilted optical axis. The substrate can be visualized as a “tilted wire medium” and is treated in our model as a continuous medium with no granularity.

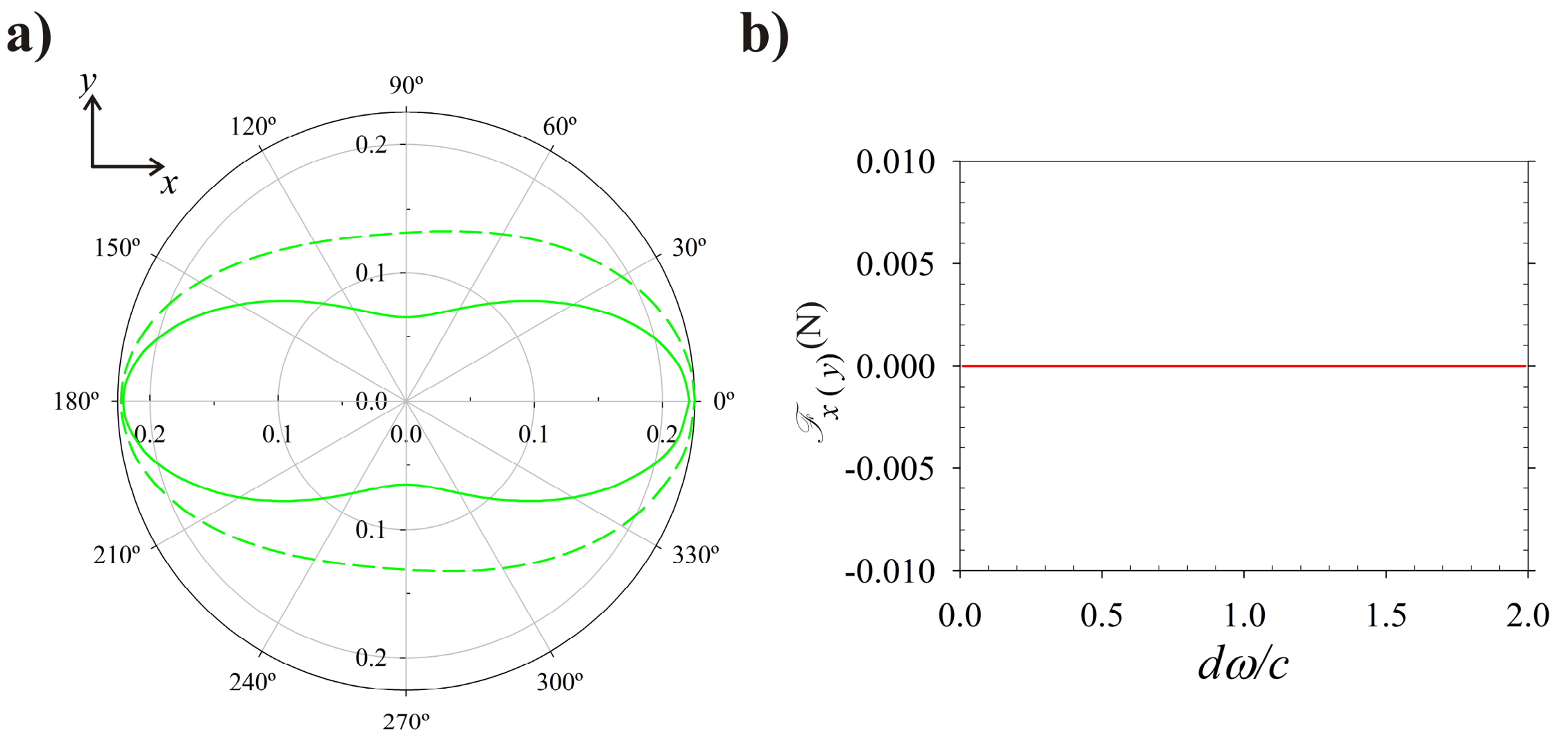

**Fig. 2**. **(a)** Polar plot of the intensity of the field back-scattered by the substrate (arbitrary normalization) calculated at a radial distance of $5\lambda_0$ from the vertical dipole. Dashed line: total field ($\left|\mathbf{E}^s\right|^2$) intensity. Solid line: intensity of the *z*-component ($\left|\mathbf{E}_z^s\right|^2$) of the electric field. The substrate is a uniaxial dielectric half-space ($\varepsilon_{\|}=1$ and $\varepsilon_{\alpha\alpha}=-10^6$) with tilted optical axes ($\alpha_0=45^{\circ}$) at the normalized distance is $d\omega/c=1$. **(b)** Plot of the lateral force $\mathscr{F}_i^s$ ($i=\{x,y\}$) as a function of $d\omega/c$ for the same configuration as in (a).

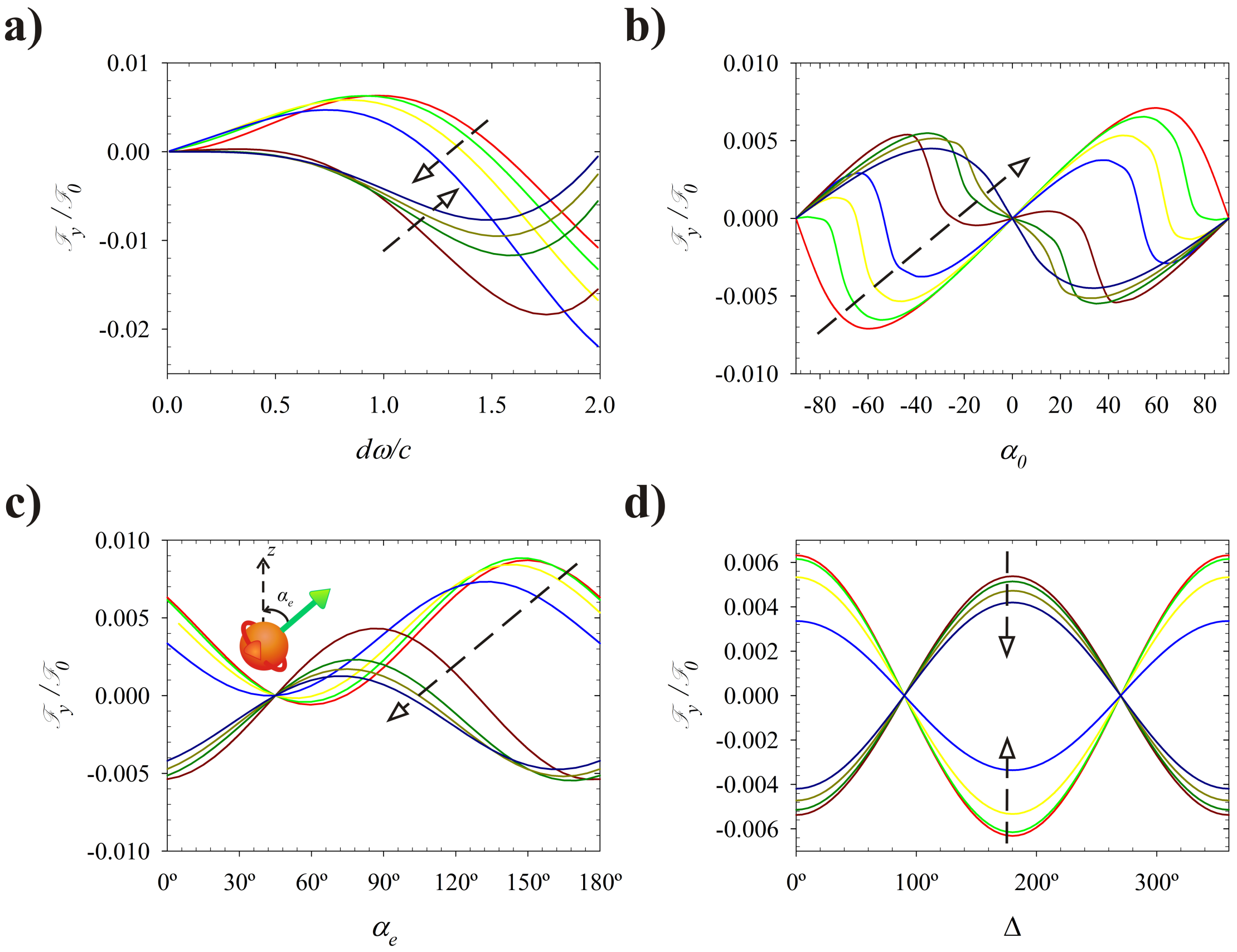

**Fig. 3**. Plot of the (*y*-component of the) lateral force $\mathscr{F}_y^s$ for a uniaxial material with $\varepsilon_{\|}=1$ for $\varepsilon_{\alpha\alpha}=-10^6,-10,-4,-2,-0.5,-0.2,-0.1,-0.01$ **(a)** $\mathscr{F}_y^s$ as a function of $d\omega/c$ for $\alpha_0=45^{\circ}$, $\alpha_m=\alpha_e=0$ and $\Delta=0^{\circ}$. **(b)** $\mathscr{F}_y^s$ as function of the optical axis tilt angle $\alpha_0$ for $d\omega/c=1$, $\alpha_m=\alpha_e=0$ and $\Delta=0^{\circ}$. **(c)** $\mathscr{F}_y^s$ as a function of $\alpha_e$ with $\alpha_0=45^{\circ}$, $d\omega/c=1$, $\alpha_m=\alpha_e$ and $\Delta=0^{\circ}$. **(d)** $\mathscr{F}_y^s$ as a function of $\Delta$ with $\alpha_0=45^{\circ}$, $d\omega/c=1$ and $\alpha_m=\alpha_e=0$. In all panels the arrow indicates the direction of increasing $\varepsilon_{\alpha\alpha}$ (from the red color, for $\varepsilon_{\alpha\alpha}=-10^6$, towards the dark blue color, for $\varepsilon_{\alpha\alpha}=-0.01$).

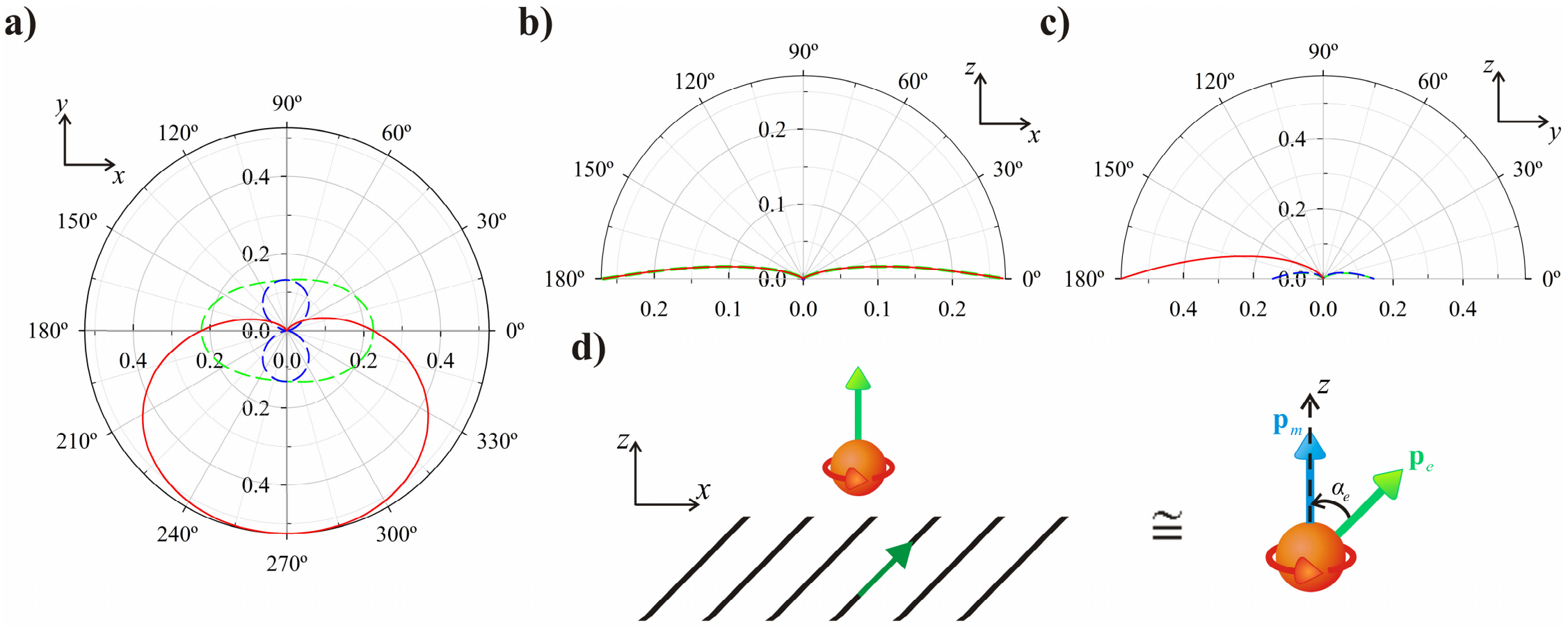

**Fig. 4**. **(a)** Polar plot of the scattered field intensity ($|\mathbf{E}^s|^2$) (arbitrary normalization) in the $z = d$ plane calculated at the $5\lambda_0$ - distance from the particle for $\alpha_0 = 45^\circ$ at the normalized distance $d\omega / c = 1$ for i) a single vertical electric dipole (green color), ii) single vertical magnetic dipole (blue color) and iii) vertical collinear electric and magnetic dipoles (red color). The amplitudes of the electric and magnetic dipoles are related as $p_e = (1/\eta_0)\, p_m$ . **(b)** Same as (a) in the $y = 0$ plane. **(c)** Same as (a) in the $x = 0$ plane. **(d)** Two collinear electric and magnetic dipoles standing above a uniaxial substrate with tilted optical axes behave effectively as two non-collinear electric and magnetic dipoles (Huygens source), leading to an asymmetric radiation pattern and a lateral recoil force.

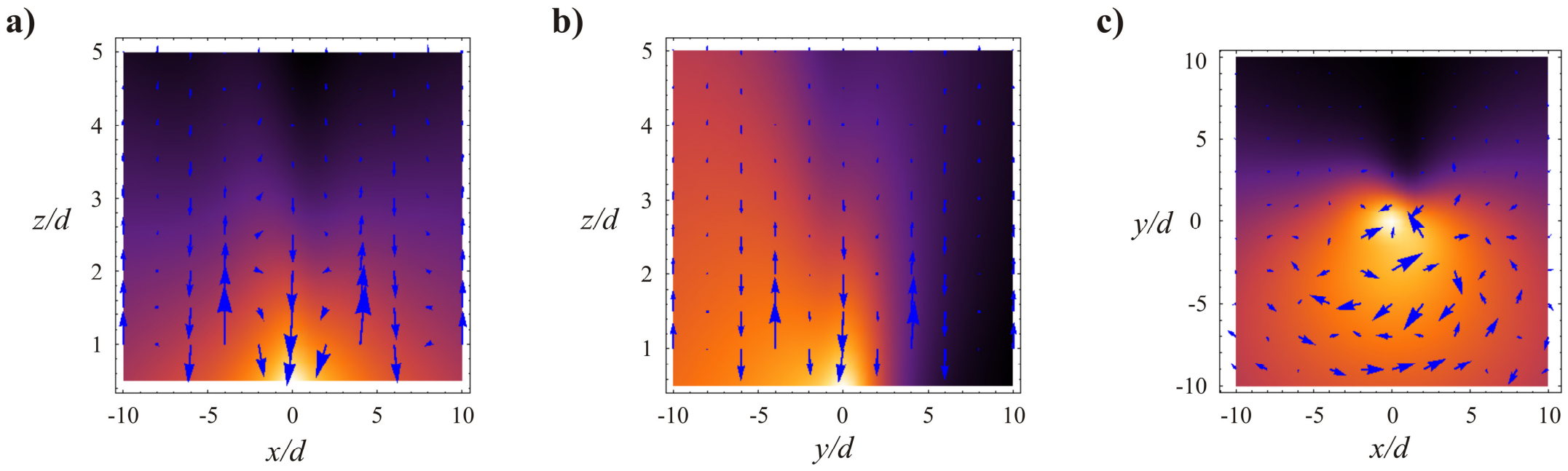

**Fig. 5**. **(a)** Density plot of the scattered field intensity ($|\mathbf{E}^s|^2$) (in arbitrary units) at the *xoz* plane for a composite particle with $\mathbf{p}_e = (1/\eta_0)\, p_m \hat{\mathbf{z}}$ and $\mathbf{p}_m = p_m \hat{\mathbf{z}}$. The tilt angle is $\alpha_0 = 45^{\mathrm{o}}$ and $d\omega / c = 1$. The arrows show a time-snapshot (*t=0*) of the in-plane electric field back-scattered by the substrate. **(b)** Same as (a) but in the *yoz* plane. **(c)** Same as (a) but in the $z = d$ plane.